\documentclass[preprint,12pt]{elsarticle}

\usepackage{amssymb}
\usepackage{amsmath}
\usepackage{booktabs}
\usepackage{multirow}
\usepackage{url}
\usepackage{float}
\usepackage{tabularx}
\usepackage{multirow}
\usepackage{caption}
\usepackage{subcaption}
\usepackage{xcolor}
\usepackage{verbatim}
\usepackage{hyperref} 
\journal{IJEPES}

\begin{document}

\begin{frontmatter}

%% Title, authors and addresses

%% use the tnoteref command within \title for footnotes;
%% use the tnotetext command for theassociated footnote;
%% use the fnref command within \author or \address for footnotes;
%% use the fntext command for theassociated footnote;
%% use the corref command within \author for corresponding author footnotes;
%% use the cortext command for theassociated footnote;
%% use the ead command for the email address,
%% and the form \ead[url] for the home page:
%% \title{Title\tnoteref{label1}}
%% \tnotetext[label1]{}
%% \author{Name\corref{cor1}\fnref{label2}}
%% \ead{email address}
%% \ead[url]{home page}
%% \fntext[label2]{}
%% \cortext[cor1]{}
%% \affiliation{organization={},
%%             addressline={},
%%             city={},
%%             postcode={},
%%             state={},
%%             country={}}
%% \fntext[label3]{}

\title{ Novel Quality Measure and Efficient Resolution of Convex Hull Pricing for Unit Commitment }

%% use optional labels to link authors explicitly to addresses:
%% \author[label1,label2]{}
%% \affiliation[label1]{organization={},
%%             addressline={},
%%             city={},
%%             postcode={},
%%             state={},
%%             country={}}
%%
%% \affiliation[label2]{organization={},
%%             addressline={},
%%             city={},
%%             postcode={},
%%             state={},
%%             country={}}

\author[inst1]{Mikhail A. Bragin}%\fnref{1}}
\author[inst2]{Farhan Hyder\corref{cor1}}
\author[inst2]{Bing Yan}
\author[inst1]{Peter B. Luh\fnref{2}}
\author[inst4]{Jinye Zhao}
\author[inst4]{Feng Zhao}
\author[inst4]{Dane A. Schiro}
\author[inst4]{Tongxin Zheng}

% \affiliation[inst1]{organization={Department of Electrical and Computer Engineering, University of California},%Department and Organization
%             %addressline={Address Two}, 
%             city={Riverside},
%             postcode={92521}, 
%             state={CA},
%             country={USA}}
%             \fntext[1]{At the time of writing of the paper, M. A. Bragin was with the Department of Electrical and Computer Engineering, University of Connecticut, Storrs, CT 06269, US.}

\affiliation[inst1]{organization={Department of Electrical and Computer Engineering, University of Connecticut},%Department and Organization
            %addressline={Address Two}, 
            city={Storrs},
            postcode={06269}, 
            state={CT},
            country={USA}} 
            \fntext[2]{P. B. Luh, who was the co-supervisor of this project, tragically passed away in November 2022. He was a professor emeritus of the Department of Electrical and Computer Engineering, University of Connecticut, Storrs, CT 06269, US, and with the Department of Electrical Engineering, National Taiwan University, Taipei 10617, Taiwan. As a tribute to our dear friend and mentor, the remaining coauthors dedicate this paper to commemorating Dr. Luh's contributions and the legacy.}

            \affiliation[inst2]{organization={Electrical and Microelectronic Engineering, Rochester Institute of Technology},%Department and Organization
            %addressline={Address One}, 
            city={Rochester},
            postcode={14623}, 
            state={NY},
            country={USA}}
            
\affiliation[inst4]{organization={Advanced Technology and Solutions, ISO New England},%Department and Organization
            %addressline={Address Two}, 
            city={Holyoke},
            postcode={01040}, 
            state={MA},
            country={USA}}
            
\cortext[cor1]{Corresponding author, E-mail address: fh6772@rit.edu (F. Hyder).}

%Tracking in ON for everyone

\begin{abstract}
%% Text of abstract
Electricity prices determined by economic dispatch that do not consider fixed costs may lead to significant uplift payments. However, when fixed costs are included, prices become non-monotonic with respect to demand, which can adversely impact market transparency. To overcome this issue, convex hull (CH) pricing has been introduced for unit commitment with fixed costs. Several CH pricing methods have been presented, and a feasible cost has been used as a quality measure for the CH price. However, obtaining a feasible cost requires a computationally intensive optimization procedure, and the associated duality gap may not provide an accurate quality measure. This paper presents a new approach for quantifying the quality of the CH price by establishing an upper bound on the optimal dual value. The proposed approach uses Surrogate Lagrangian Relaxation (SLR) to efficiently obtain near-optimal CH prices, while the upper bound decreases rapidly due to the convergence of SLR. Testing results on the IEEE 118-bus system demonstrate that the novel quality measure is more accurate than the measure provided by a feasible cost, indicating the high quality of the upper bound and the efficiency of SLR.

\end{abstract}

\begin{keyword}
%% keywords here, in the form: keyword \sep keyword
Electricity Markets \sep Convex Hull Pricing
\sep Surrogate Lagrangian Relaxation  %\sep ramp-rate constraints
%% PACS codes here, in the form: \PACS code \sep code
%\PACS 0000 \sep 1111
%% MSC codes here, in the form: \MSC code \sep code
%% or \MSC[2008] code \sep code (2000 is the default)
%\MSC 0000 \sep 1111
\end{keyword}

\end{frontmatter}

%% \linenumbers

%% main text
\section{Introduction}
\label{sec:1}
In the electricity markets of the United States, prices are usually determined based on the economic dispatch (ED) problem \cite{litvinov2019electricity}. However, since commitment costs such as start-up and no-load costs are not factored in, uplift payments\footnote{Uplift payments are additional payments made to generators to cover certain costs that may not be included in the market price, such as start-up and shutdown costs.} may become substantial. When prices are determined based on the unit commitment (UC), which does account for such costs, the associated binary commitment-related variables lead to non-convexity. Consequently, prices based on a UC problem are non-monotonic with respect to demand \cite{gribik2007market}, thereby reducing market transparency.

Convex hull (CH) pricing has been introduced as a possible resolution to the issues associated with fixed costs, binary variables, uplift payments, and non-monotonic prices based on the unit commitment (UC) problem  \cite{gribik2007market}. CH pricing continues to be an active area of research today \cite{andrianesis2021computation, stevens2022application}. One approach to obtaining CH prices is to solve the dual problem of the UC problem, with optimal CH prices representing optimal Lagrangian multipliers (\(\lambda^*\)) \cite{wang2013subgradient,wang2013extreme}. The difference between a feasible cost to the UC problem and a dual value yields the uplift payment. However, this approach can face challenges related to the convergence of Lagrangian multipliers by using standard subgradient methods associated with high computational effort as well as multiplier zigzagging \cite{luh1998algorithm}. Various methods have been proposed to solve the CH pricing problem, with the quality of the resulting CH prices typically measured using a feasible cost that quantifies their proximity to the optimal prices, i.e., duality gap \cite{yu2020convex}.

Challenges related to the current quality-measuring approach include 1. the substantial computational effort needed to obtain a feasible cost, and 2. a potentially insufficiently tight duality gap to provide a high-quality measure, as illustrated schematically in Figure \ref{fig_1} below. Consequently, a more robust alternative measure (upper bound) is desired. The innovative CH price quality measure is defined as the difference between the upper and lower bounds of the optimal dual value. While the lower bound can be derived from the best available dual value, a novel approach to determining the upper bound is necessary.

%\textcolor{red}{(Can I get the source figure for fig.1 to improve the quality of the image?)}
\begin{figure}[h!]
\centering
\includegraphics[scale = 0.9]{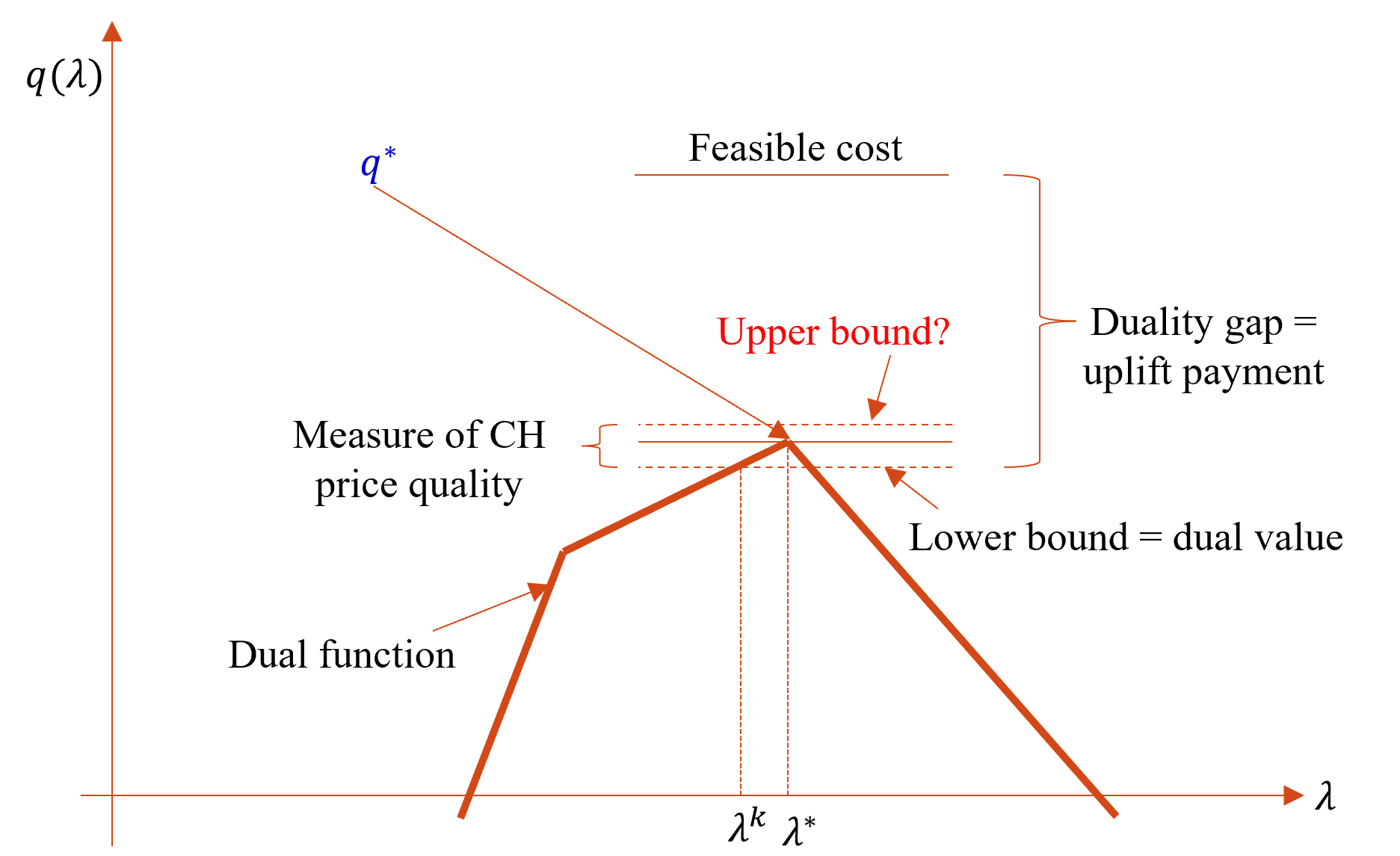}
\caption{A schematic demonstration of a dual function, feasible cost, standard duality gap, and the novel proposed measure of CH price-quality.}
\label{fig_1}
\end{figure}

% The goal of this paper is to provide a high-quality measure of CH prices through the development of a novel upper bound to the optimal dual value, and to efficiently obtain CH prices.

% After briefly presenting the UC problem in Section II, a novel CH-price quality measure is developed in Section III. It is based on the difference between the innovative upper bound \(\Bar{q}\) and the best available dual value (\(q(\lambda^k)\)) as shown in Figure 1. The near-optimal CH prices are obtained by using the Surrogate Lagrangian Relaxation (SLR) method \cite{bragin2015convergence} which overcame the difficulties of standard sub-gradient methods.  Meanwhile, the fast convergence of SLR also leads to a fast decrease of the upper bound. Testing results based on the IEEE 118-bus system indicate that the CH-price quality measure is much more accurate than the measure that a feasible cost can provide - demonstrating both the high quality of the upper bound and the efficiency of the SLR method. 

This paper aims to provide a high-quality measure of CH prices through the development of a novel upper bound to the optimal dual value and the efficient computation of CH prices. We first present the UC problem in Section \ref{sec2} and then develop a CH-price quality measure based on the difference between the innovative upper bound \(\Bar{q}\) and the best available dual value (\(q(\lambda^k)\)) in Section \ref{sec:3}. The near-optimal CH prices are obtained using the Surrogate Lagrangian Relaxation (SLR) method \cite{bragin2015convergence}, which overcomes the challenges of standard sub-gradient methods and offers fast convergence of multipliers as well as the upper bound to the optimal dual value. Testing results based on the IEEE 118-bus system demonstrate the superiority of the proposed CH-price quality measure and the efficiency of the SLR method.

\section{UC Problem Formulation for Convex Hull Pricing }
\label{sec2}
This section presents the formulation of the UC problem, which serves as the foundation for the development of a novel CH-price quality measure and further presentation of CH prices. The section is divided into two subsections: subsection \ref{21} focuses on the formulation without transmission constraints, which are frequently omitted in existing literature for the calculation of CH prices, and subsection \ref{22} includes transmission constraints for a more realistic representation of the power system. 

\subsection{Unit Commitment Formulation without Transmission Constraints}  \label{21}
Consider a UC problem with \(I\) units indexed by \(i\) and \(T\) hours indexed by \(t\). 
\begin{enumerate}
    \item Objective Function: The goal is to minimize the total cost as:
        \begin{equation}
        \begin{aligned}
        \min_{\left\{x_i,u_i,p_i,r_i\right\}}{\left\{\sum_{i=1}^{I}{z_i\left(x_i,u_i,p_i,r_i\right)}\right\}},
        %\underset{G=[{{g}_{it}}]}{\mathop{\text{Min}}}\,\ F(G),\text{with }F(G)\equiv \sum\limits_{i=1}^{N}{\sum\limits_{t=1}^{T}{{{f}_{it}}({{g}_{it}})}},
        \end{aligned}
        \end{equation}
        
        where \(z_i\left(x_i,u_i,p_i,r_i\right)= \displaystyle \sum_{t=1}^{T}\left(c_i^E\left(p_{i,t}\right)+c_i^S\cdot u_{i,t}+c_i^N\cdot x_{i,t}\right)\) represents the total cost of unit \(i\), including energy costs, 
%        reserve costs, 
        start-up costs, and no-load costs. Decision variables at time \(t\) are: 1. binary commitment \(x_{i,t}\) and start-up \(u_{i,t}\); and 2. continuous dispatch \(p_{i,t}\) 
        %and reserve \(r_{i,j,t}\) of type \(j\)
        .
    
    \item System-Wide Constraints

%    \begin{enumerate}
%        \item Reserve Requirements: We consider three types of reserve (regulation, regulation plus spinning, and operating reserve) indexed by \(j\) = 1, 2, and 3, respectively, following \cite{nerc}:
%            \begin{equation}
%            \begin{aligned}
%            	\sum_{i=1}^{I}r_{i,j,t}\geq R_{j,t},j=1,\ldots,3,t=1,\ldots,T,	\label{eq2}		
%            \end{aligned}
%            \end{equation}
%            where \(R_{j,t}\) represent reserve requirements. 

%        \item 
        System Demand: The total power generation equals the demand at each time \(t\): 
            \begin{equation}
            \begin{aligned}
            	\sum_{i=1}^{I}p_{i,t}=D_t,t=1,\ldots,T,	\label{eq3}	
             \end{aligned}
            \end{equation}
            where \(p_{i,t}\) are generation levels and \(D_t\) is the demand at time \(t\).
%        \end{enumerate}
  
    \item Unit-Level Constraints 
        \begin{enumerate}
          \item Generation Capacity Constraints: The generation capacity for online units $(x_{i,t} = 1)$ is restricted by a lower limit and an upper limit, denoted by $P_i^{min}$ and $P_i^{max}$, respectively. Meanwhile, when a unit is offline, represented by the corresponding commitment status $x_{i,t} = 0$, 
          %both its generation and reserve levels are zero. 
          its generation level is zero. 
          The mathematical constraints capturing these relationships are provided below:
            \begin{equation}
            \begin{aligned}
            x_{i,t}\cdot P_i^{min} \le p_{i,t} \le x_{i,t}\cdot P_i^{max},	\label{eq4}	
            \end{aligned}
            \end{equation}
          %The regulation reserve ($j=1$) is subtracted to ensure that there is enough available generation capacity to meet the demand even in unexpected situations, such as a sudden increase in demand or unexpected generator outages. By subtracting the reserve from the available generation capacity of an online unit $(x_{i,t} = 1)$, the system ensures that there is always enough generation capacity available to meet the demand, even in the worst-case scenarios. This approach helps to ensure the reliability and stability of the power system \textcolor{blue}{(Bing, can you please check whether my explanation is ok)}:  

%            \begin{equation}
%            \begin{aligned}
%            x_{i,t}\cdot P_i^{min} \le p_{i,t}-r_{i,1,t},	\label{eq5}	
%            \end{aligned}
%            \end{equation}
%          Reserve levels are also upper bounded by maximum allowed levels as: 
%            \begin{equation}
%           \begin{aligned}
%            0\le r_{i,j,t} \le r_{i,j}^{max}.	 \label{eq6}	
%            \end{aligned}
%           \end{equation}
            % where \(x_{i,t}\) are binary commitment decision variables.

        \item Start-Up Constraints: 
            Start-ups are captured by using binary variables \(u_{i,t}\) as: 
             \begin{equation}
             \begin{aligned}
                x_{i,t}-x_{i,t-1}\le u_{i,t}.	\label{eq7}
             \end{aligned}
             \end{equation}
          In the above relationship, a start-up cost is incurred when $u_{i,t}=1,$ which is only possible when $x_{i,t} = 1$ and $x_{i,t-1}=0.$
        \item  Minimum Up- and Down-Time Constraints: Unit \(i\) stays online (offline) for its minimum up- (down-) time \(l_i (L_i)\):
             \begin{equation}
             \begin{aligned}
                \sum_{j=t-L_i+1}^{t}u_{i,j}\le x_{i,t},\ \sum_{j=t-l_i+1}^{t}u_{i,j}\le1-x_{i,t-l_i}.	\label{eq8}
             \end{aligned}
             \end{equation}

        \item Ramp-Rate Constraints: The change of generation levels between two consecutive hours cannot exceed its ramp-rate requirements:
             \begin{equation}
             \begin{aligned}
                p_{i,t}-p_{i,t-1}\le R_i\cdot x_{i,t-1}+V_i\cdot\left(1-x_{i,t-1}\right),	\label{eq9}
             \end{aligned}
             \end{equation}
             
             \begin{equation}
             \begin{aligned}
                p_{i,t-1}-p_{i,t}\le R_i\cdot x_{i,t}+V_i\cdot\left(1-x_{i,t}\right),	\label{eq10}
             \end{aligned}
             \end{equation}		
                where \(R_i\) are ramp-up/down and \(V_i\) is the start-up/shut-down ramp rates. 

      \end{enumerate}
\end{enumerate}
\subsection{Unit Commitment Formulation with Transmission Constraints} \label{22}
The above UC problem formulation does not consider transmission constraints. However, transmission constraints are often included in practical situations. In such cases, the following constraints are added to the formulation instead of the demand constraint in \eqref{eq3}:

\begin{enumerate}
\item DC Power Flow Constraints: The power flow on each transmission line \(l\) at time \(t\) is given by:
\begin{equation}
\begin{aligned}
f_{l,t}=\frac{\theta_{s\left(l\right),t}-\theta_{r\left(l\right),t}}{X_l},l=1,\ldots,L,t=1,\ldots,T. \label{eq11}
\end{aligned}
\end{equation}
Here \(\theta_{s\left(l\right),t}\) and \(\theta_{r\left(l\right),t}\) are phase angles at sending and received buses of line \(l\), and \(X_l\) is line \(l\) impedance.

\item Nodal Flow Balance Constraints: Nodal flow balance is maintained at each node \(n\)=1,\(\ldots,N\) by the following equation:
    \begin{equation}
    \begin{aligned}
        \sum_{l:r\left(l\right)=n} f_{l,t}+p_{n,t}=\sum_{l:s\left(l\right)=n} f_{l,t}+D_{n,t},\ t=1,\ldots,T.	\label{eq12}
    \end{aligned}
    \end{equation}

\item Transmission Capacity Constraints: The power flow on each transmission line \(l\) is limited by its minimum and maximum capacity:
    \begin{equation}
    \begin{aligned}
        f_l^{min}\le f_{l,t}\le f_l^{max},l=1,\ldots,L,t=1,\ldots,T. \label{eq13}	
    \end{aligned}
    \end{equation}
\end{enumerate}

\section{Convex Hull Pricing through the Dual Problem} 
\label{sec:3}
This section is on the development of a novel measure to quantify CH-price quality. After briefly presenting a dual problem in subsection \ref{31} as well as the SLR method in subsection \ref{32}, the novel measure is developed in subsection \ref{33}. 

\subsection{The Dual Problem of the UC Problem} \label{31}
For simplicity, the explanation will rest upon the formulation with system demand constraints; a formulation with transmission constraint follows analogously. %\textcolor{blue}{(If other CH papers do not consider transmission constraints, we need to state this as a contribution.) - Some papers do consider transmission constraints} 
The dual problem to the UC problem is defined as \cite{hua2016convex}:
\begin{equation}
\begin{aligned}
\max_{\lambda} q(\lambda), \label{eq14}
 \end{aligned}
 \end{equation}
 
\noindent where \(q(\lambda)\)  is the dual function defined as:

\begin{equation}
\begin{aligned}
q\left(\lambda\right)=\min_{\left\{p,x,u\right\}\in\mathfrak{F}}{L\left(\lambda,p,x,u\right)}. \label{eq15}
 \end{aligned}
 \end{equation}

\noindent Here \(\mathfrak{F}\) is a feasible set delineated by constraints \eqref{eq4}-\eqref{eq10}, the minimization within \eqref{eq15} is the ``relaxed problem,” and \(L\left(\lambda,p,x,u\right)\) is the Lagrangian function defined as: 

\begin{equation}
\begin{aligned}
L\left(\lambda,p,x,u\right)=\sum_{i=1}^{I}{z_i\left(x_i,u_i,p_i\right)}+\lambda\cdot g(p), \label{eq16}
 \end{aligned}
 \end{equation}

 \noindent where \(\lambda=\left\{\lambda_t\right\}\) is a vector of Lagrangian multipliers obtained after relaxing 
 %reserve \eqref{eq2} 
 demand constraints \eqref{eq3}. The corresponding constraint violations are collectively denoted as \(g(p)\) for the compactness of notation. The CH prices are optimal multipliers \(\lambda^\ast\) of the dual problem \eqref{eq14} \cite{gribik2007market}. 

 The methodology to be presented will be based on the above. Extension for the formulation with transmission follows the same logic with the exception that nodal flow balance constants \eqref{eq12} are relaxed rather than system demand constraints \eqref{eq3}, in which case the multipliers \(\lambda
 % =\left\{\lambda_{i,t}\right\}
 \) will be a matrix.  

\subsection{Surrogate Lagrangian Relaxation Method \cite{bragin2015convergence}}  \label{32}
 In this subsection, our recent Surrogate Lagrangian Relaxation (SLR) Method will be briefly introduced. Within the method, multipliers are updated based on “surrogate” subgradients \(g\left(\tilde{p}^k\right)\) (which are levels of constraint violations of relaxed constraints), and stepsizes \(s_{SLR}^k\) as: 
\begin{equation}
\begin{aligned}
\lambda_t^{k+1}=\lambda_t^k+s_{SLR}^k\cdot g\left(\tilde{p}^k\right), \label{eq17}
\end{aligned}
\end{equation}
where, 
\begin{equation}
\begin{aligned}
s_{SLR}^k=\left(1-\frac{1}{M\cdot k^{1-\frac{1}{k^\rho}}}\right) \cdot s_{SLR}^{k-1} \cdot \frac{|| g(\tilde{p}^{k-1}) ||}{|| g(\tilde{p}^k) ||}, M>1, 0<\rho<1. \label{eq18}
\end{aligned}
\end{equation}
Here, “tilde” (\(\texttildelow\)) indicates that optimization of the relaxed problem is subject to the following “surrogate optimality condition” \cite[p. 178, eq. (12)]{bragin2015convergence}: $L\left(\lambda^k,\ \tilde{x}^k,\ \tilde{u}^k,\ \tilde{p}^k\right)<$ 
$L\left(\lambda^k,\ \tilde{x}^{k-1},\ \tilde{u}^{k-1},\ \tilde{p}^{k-1}\right)$, rather than to full minimization the Lagrangian function in \eqref{eq15}, with $L\left(\lambda^k,\ \tilde{x}^k,\ \tilde{u}^k,\ \tilde{p}^k\right)$ being a ``surrogate dual value” (a value that the Lagrangian function \eqref{eq16} attains at solutions $\tilde{x}^k,\ \tilde{u}^k,\ \tilde{p}^k$). The primary advantages of the SLR method lie in the fact that it does not require solving all sub-problems simultaneously and only needs to satisfy the surrogate optimality condition mentioned earlier. Both of these factors contribute to a significant reduction in computational effort. Moreover, surrogate directions do not change drastically from one iteration to the next, and zigzagging difficulties are thus alleviated. Thereby also contributing to the overall significant reduction of the computational effort. 
%The above condition significantly reduces the computational effort and alleviates the zigzagging of multipliers.
% thereby reducing the number of iterations required for convergence. 

\subsection{Novel Quality Measure for Convex Hull Pricing }  \label{33}
This subsection is on a novel way to obtain the upper bound \(\bar{q}\) to the optimal dual value \(q\left(\lambda^\ast\right)\), approaching \(q\left(\lambda^\ast\right)\) from above. Together with dual values \(q\left(\lambda^k\right)\) (which are a lower bound), the novel CH-price quality measure will be obtained as a relative difference between \(\bar{q}\) and \(q\left(\lambda^k\right)\). 

As proved within the “Surrogate” Subgradient Method (SSM) \cite[pp. 704-706]{Zhao1999surrogate}, if the following condition on stepsizes holds
\begin{equation}
\begin{aligned}
s^k < \frac{q^* - \Tilde{L} (\lambda^k,x^k,u^k,p^k) }{|| \Tilde{g}(p^k)||^2}, \label{eq19}
\end{aligned}
\end{equation}
then multipliers approach $\lambda^*$ at every iteration:
\begin{equation}
\begin{aligned}
||\lambda^* -  \lambda^{k+1}||^2 < ||\lambda^* - \lambda^k||^2. \label{eq20}
\end{aligned}
\end{equation}
However, the optimal dual value $q^*$ required to set stepsizes is generally not known for any given problem. Therefore, when stepsizes are set by using any other rule without using $q^*$, then multipliers may not approach $\lambda^*$ (or any other point) at every iteration. This ``non-approaching'' property will be exploited to derive the novel quality measure. Specifically, if one can detect that multipliers do not approach $\lambda^*$, then this is only because stepsizes are ``too large,'' that is, stepsizes violate \eqref{eq19}. This result is summarized in the following Lemma.  
% In the context of this paper, divergence is understood as not converging to $\lambda^*$ or any common limit within a finite number of iterations. 

% The main premise behind the derivation of the upper bound is the fact proved within the “Surrogate” Subgradient Method (SSM) \cite[pp. 704-706]{Zhao1999surrogate} stating that if the following condition on stepsizes holds 

% \begin{equation}
% \begin{aligned}
% s_{SSM}^k < \frac{q^* - \Tilde{L} (\lambda^k,x^k,u^k,p^k,r^k) }{|| \Tilde{g}(p^k,r^k)||^2}.
% \end{aligned}
% \end{equation}

% Multipliers are then updated using (17) with stepsizes satisfying (19) approach \(\lambda^\ast\) in the following way:  \footnote{If optimal multipliers are not unique, then \(||\lambda^* – \lambda^k||\), which is the Euclidean distance to \(\lambda^\ast\), is understood as the distance to a set \(\Lambda^\ast\equiv{\lambda|q\left(\lambda\right)=q(\lambda^\ast)}\). }

% The result opposite to the above is stated in the Lemma below. \\
\textbf{Lemma}: If there exists \(\kappa\) so that \(\lambda^\ast\) move away from \(\lambda^\ast\), i.e.,

\begin{equation}
\begin{aligned}
||\lambda^* -  \lambda^{k+1}||^2 \geq ||\lambda^* - \lambda^k||^2,
\end{aligned} \label{eq21}
\end{equation}
then
\begin{equation}
\begin{aligned}
s^\kappa  \geq  \frac{q^* -  \Tilde{L} (\lambda^\kappa, x^\kappa, u^\kappa, p^\kappa)}{||\Tilde{g}(p^\kappa)||^2}.  \label{eq22}
\end{aligned}
\end{equation}

 \textbf{Proof}: Denote \eqref{eq19} as assertion A and \eqref{eq20} as B. According to \cite[pp. 704-706]{Zhao1999surrogate}, the following predicate holds true: \(\forall k:\ A\rightarrow B\).  In the proof, the following relation will be helpful: 
\begin{equation}
\begin{aligned}
\lnot\left(A\rightarrow B\right)=\lnot A\vee B. \label{eq23}
\end{aligned}
\end{equation}

Negation of the predicate \(\forall k:\ A\rightarrow B\) leads to \(\exists k:\ \lnot(A\rightarrow B)\), which simplifies to \(\exists k:\ \lnot A\vee B\), which is equivalent to \(\exists k:\ B\vee\left(\lnot A\right)\) by the commutative property of disjunctions. By \eqref{eq23}, the latter predicate is equivalent to \(\exists k:(\lnot B)\rightarrow\left(\lnot A\right)\).
\qed 

According to \eqref{eq21}, there exists an overestimate of the optimal dual value \(\bar{q}>q^\ast\) such that 
\begin{equation}
\begin{aligned}
s^\kappa  \geq  \frac{\bar{q} -  \Tilde{L} (\lambda^\kappa, x^\kappa, u^\kappa, p^\kappa)}{||\Tilde{g}(p^\kappa)||^2}. \label{eq24}
\end{aligned}
\end{equation}

From \eqref{eq24}, the overestimate can be expressed as: 
\begin{equation}
\begin{aligned}
\bar{q} = s^\kappa ||\Tilde{g}(p^\kappa)||^2 + \Tilde{L}(\lambda^\kappa, x^\kappa, u^\kappa, p^\kappa). \label{eq25}
\end{aligned}
\end{equation}

However, the difficulty is that \eqref{eq25} was determined based on the inequality \eqref{eq21} which involves the unknown \(\lambda^\ast\), and the upper bound \eqref{eq25} cannot be computed. 
To resolve this difficulty, the following assumption is used.  
%\textcolor{red}{(is it lamba* or q*? If its lambda*, then can we explain why do we need it to compute eq 25 )} 

\noindent \textbf{Assumption}: Assuming that multipliers \eqref{eq17} generated by the SLR method do not approach any common value (including \(\lambda^\ast\)) during several iterations, which is a realistic assumption because stepsizes \eqref{eq17} are not set by using \eqref{eq19} with an unknown $q^*$, then conditions \eqref{eq22}, \eqref{eq24} and \eqref{eq25} hold. To determine whether multipliers approach any common value or not, the following ``multiplier divergence detection” problem \eqref{eq26} is solved where divergence is detected if \eqref{eq26} is infeasible (the multipliers diverge instead of converging):
%\textcolor{red}{(The multiplier divergence is not clear to me, they seem to be converging. I saw this in your paper level-based SLR. But the way you describe multiplier divergence over there is a little different)} 

%If (26) holds, multipliers are converging. otherwise, divergence is detected. (I will rewrite this line) 

\begin{equation}
\begin{gathered}
||\lambda - \lambda^{k+1}||^2 \leq ||\lambda - \lambda^{k}||^2, \\
||\lambda - \lambda^{k+2}||^2 \leq ||\lambda - \lambda^{k+1}||^2, \\
\cdots  \\
||\lambda - \lambda^{k+n}||^2 \leq ||\lambda - \lambda^{k+n-1}||^2. \label{eq26}
\end{gathered}
\end{equation}
\noindent With respect to \eqref{eq26}, \(\lambda\) is a vector of continuous decision variables (not to be confused with dual variables within \eqref{eq14}). 
% and $\lambda^k,\dots,\lambda^{k+n-1}$.
\qed

The above assumption is realistic. Otherwise, multipliers would always strictly approach \(\lambda^\ast\) and converge to \(\lambda^\ast\). The latter is by the design of the SLR method \cite{bragin2015convergence}, yet, the former can only be guaranteed if the SSM stepsize \eqref{eq19} is used. 

The problem \eqref{eq26} is set up starting from values $\lambda^k$ and $\lambda^{k+1}$ and one inequality. After $\lambda^{k+2}$ is obtained, the second inequality is added, and so on until a certain $k+n-1$ ($n$ is not known beforehand) iteration is reached whereby \eqref{eq26} is infeasible. 

% The problem (26) starts from \(k=1, n=1\). After multipliers are updated, \(n\) is increased by 1 until (26) admits no feasible solution. 

Since at one of the iterations $k,\dots,k+n-1$, equality \eqref{eq22} holds, the sought-for upper bound is then obtained as: 
\begin{equation}
\begin{aligned}
\Bar{q} = \max_{\kappa\in[k+1,k+n]} \{s^\kappa \cdot ‖\Tilde{g} (p^\kappa)‖^2 + \Tilde{L} (\lambda^\kappa, x^\kappa, u^\kappa, p^\kappa)\}. \label{eq27}
\end{aligned}
\end{equation}
Subsequently, values of \(k\) and \(n\) are reset as \(k\rightarrow k+n,\ n\rightarrow1\), and the procedure repeats.  

Finally, the quality measure of the CH prices is defined as  \(\frac{{\bar{q}}^k-q^k}{{\bar{q}}^k}\), where \({\bar{q}}^k\) is the best (lowest) upper bound, and \(q^k\) is the best (highest) dual value obtained up to iteration \(k\). This measure can be used as a stopping criterion. 

Unlike feasible costs, the novel upper bound approaches the optimal dual value from above since the upper bound \eqref{eq27} depends on the stepsizes approaching zero and the Lagrangian function approaching the optimal dual value, as proved in \cite{bragin2015convergence}. Another implication of the above is that the fast convergence of SLR leads to fast convergence of the upper bound to the optimal dual value.  

%%%%%%%%%%%%%%%%%%%%%%%%%%%%%%%%%%%%%%%%%%
\section{Numerical Testing}

This section is to demonstrate the performance of the novel quality measure of CH prices by comparing it with the duality gap of the problem. In Example 1, a small 2-unit UC problem is considered. In Examples 2 and 3, a UC problem based on the IEEE 118-bus system without and with transmission capacity constraints is considered. Within all examples, 24 hours are considered. The method is implemented using IBM ILOG CPLEX Optimization Studio V 12.10.0.0 on a PC with a 2.40 GHz Intel® Xeon® E-2286M CPU and 32G RAM.
% \newline
\subsection{Example 1. Small Example without transmission capacity constraints} \label{ex1}
Consider a UC problem with two units with the following characteristics: 
\begin{itemize}
    \item Unit 1: Pmin = 50 MW, Pmax = 200 MW, initial ramp rate = 150.3 MW/hr and general ramp rate = 200.6 MW/hr, generation cost: 65 \$/MW and start-up cost: 0.
    \item Unit 2: Pmin = 50 MW, Pmax = 200 MW, initial ramp rate = 70.35 MW/hr and general ramp rate = 40.7 MW/hr, generation cost: 40 \$/MW, start-up cost: \$6000.
\end{itemize}

\begin{figure}[h!]
\centering
\includegraphics[scale = 0.9]{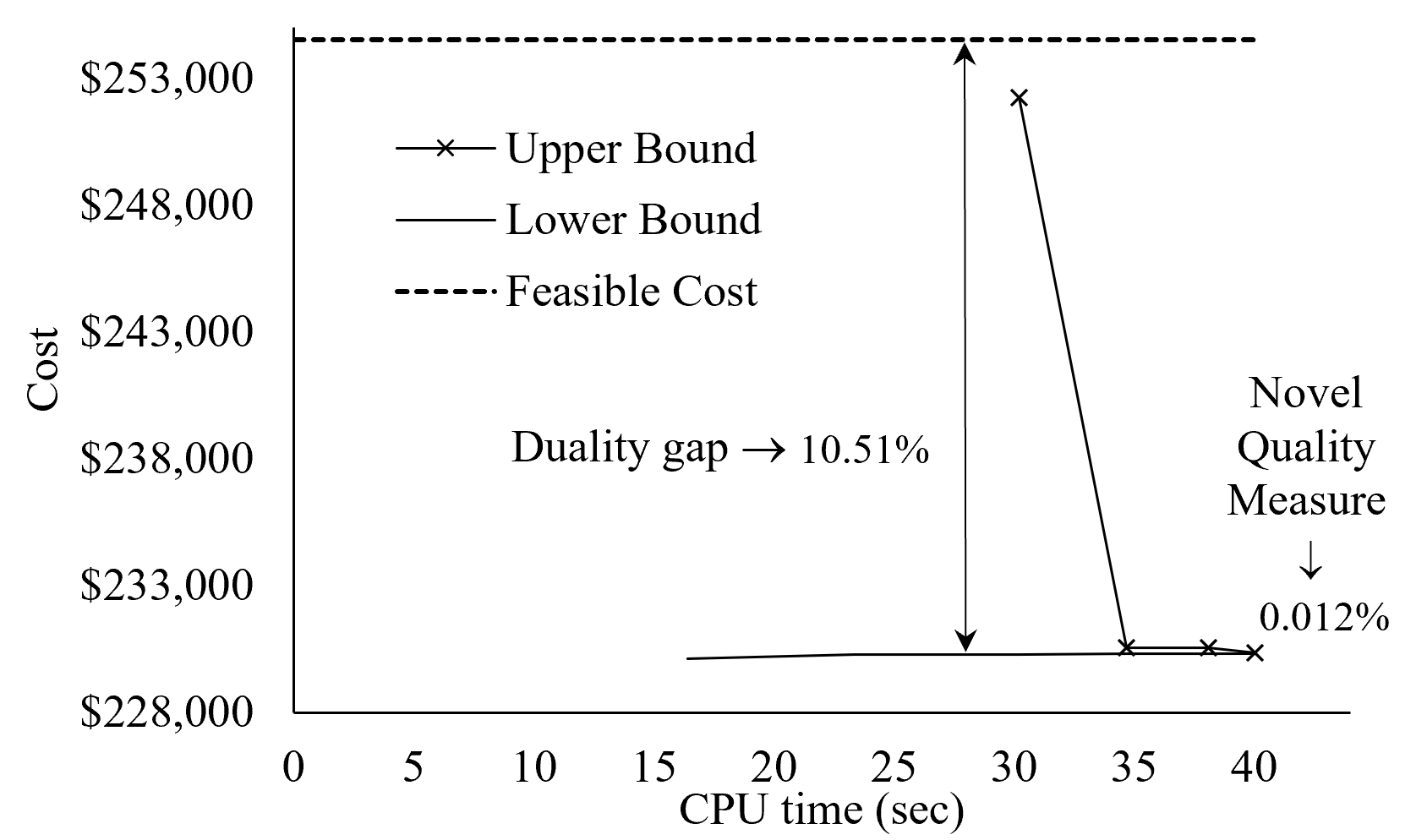}
\caption{Results for Example 1}
\label{fig_2}
\end{figure}

The results are shown in Figure \ref{fig_2}. 
Our SLR method obtains near-optimal CH prices with a quality of 0.012\% in 40 sec. Compared to the duality gap of 10.51\%, the novel quality measure of 0.012\% is significantly more accurate.
The amount of time required to obtain the upper bound is 0.448 sec, which is roughly 1\% of the total time (40 sec). The amount of time required to obtain the feasible cost is slightly longer at 0.6 sec. 

\subsection{Example 2. IEEE 118-bus Case without transmission capacity constraints}  \label{ex2}
\begin{figure}[h!]
\centering
\includegraphics[scale = 0.9]{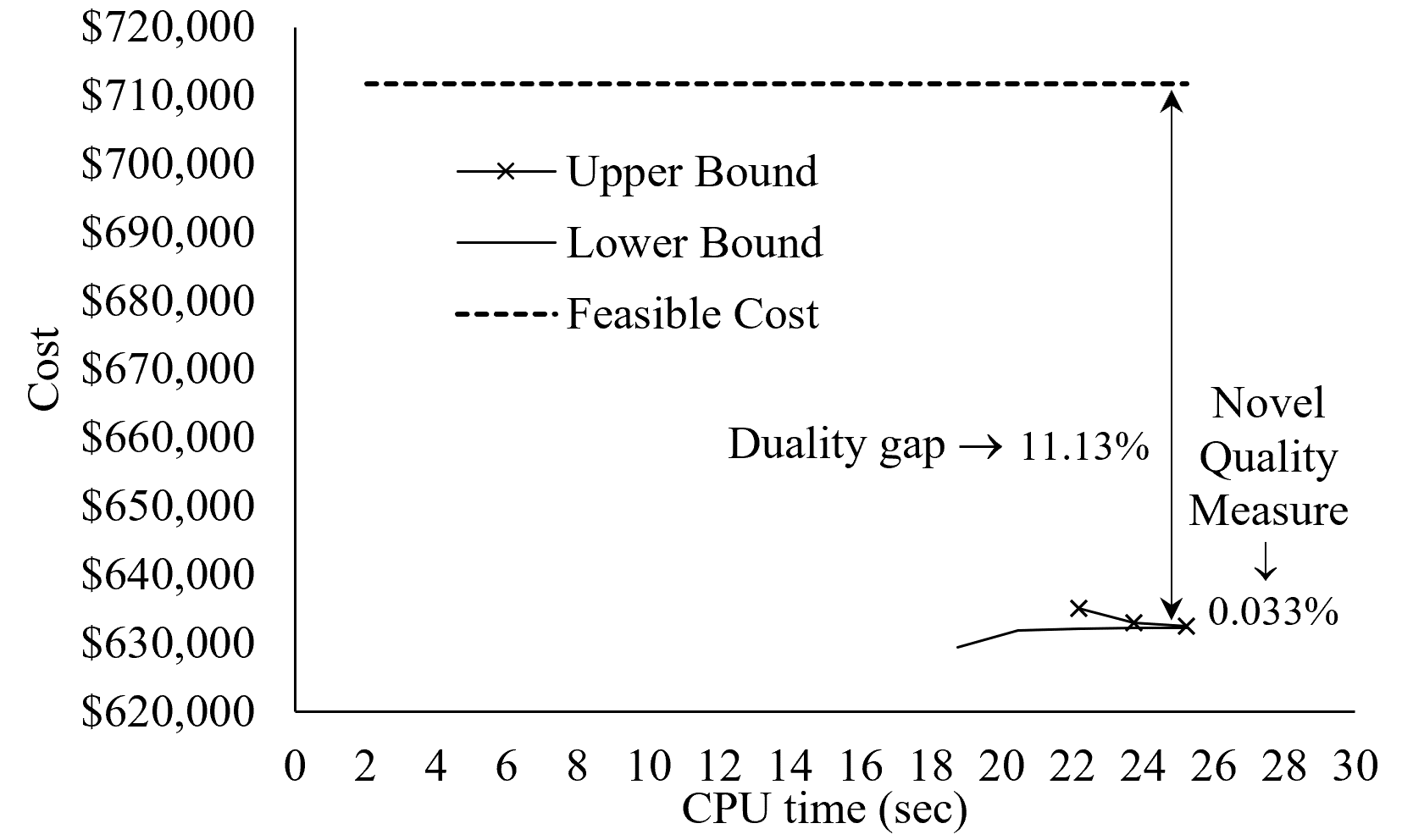}
\caption{Results for Example 2}
\label{fig_3}
\end{figure}

Consider a 54-unit, 24-hour UC problem based on the IEEE 118-bus system. The results are shown in Figure \ref{fig_3}. Compared to the duality gap of 11.13\%, the novel quality measure of 0.033\% is significantly more accurate. Near-optimal CH prices with a quality of 0.033\% are obtained in 25 sec. The amount of time required to obtain the upper bound is 0.617 sec, which is roughly 2.2\% of the total time. The amount of time required to obtain the feasible cost is slightly longer at 0.647 sec. 

\subsection{Example 3. IEEE 118-bus Case with transmission capacity constraints}  \label{ex3}
 
\begin{figure}[h!]
\centering
\includegraphics[scale = 0.9]{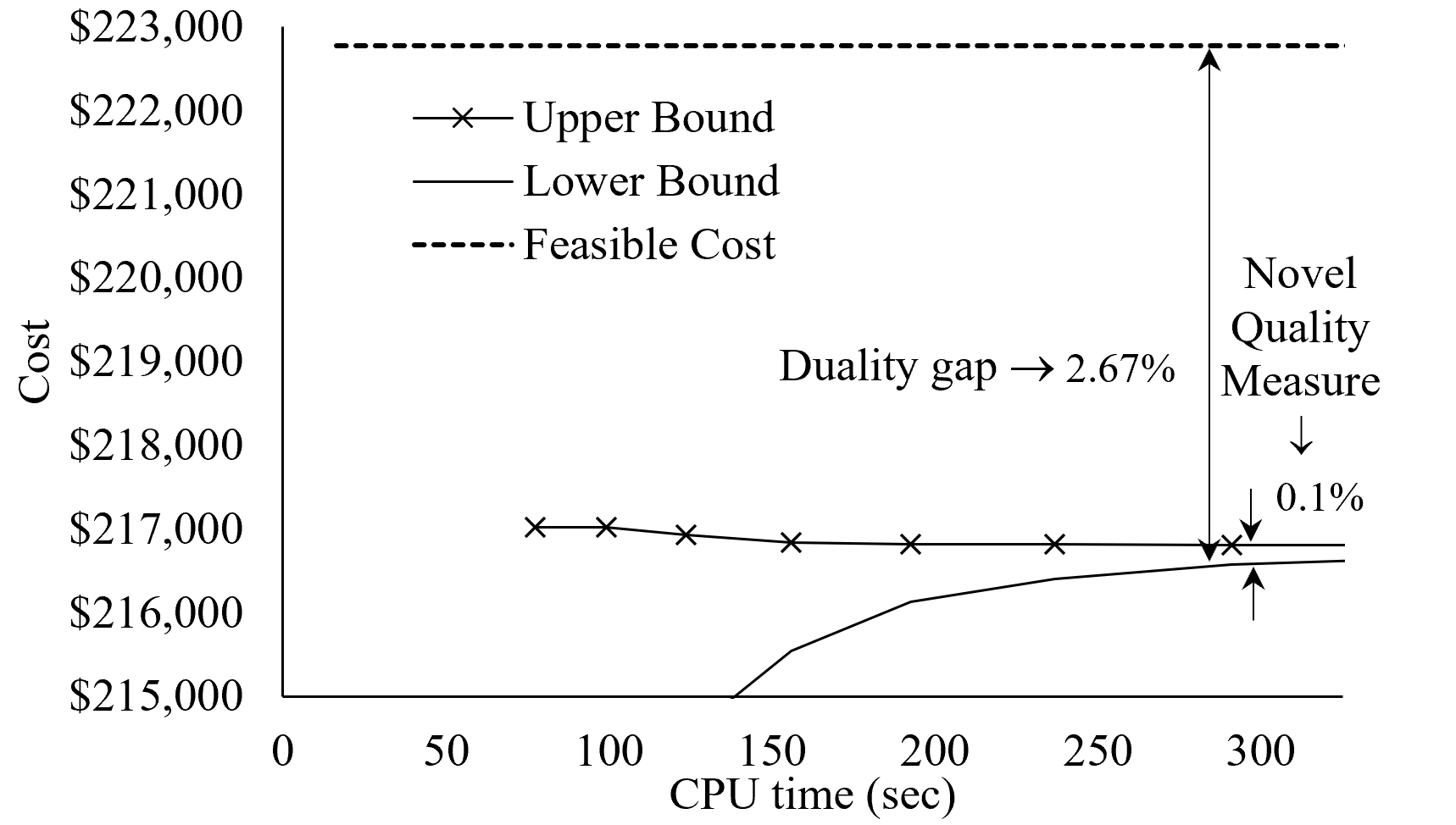}
\caption{Results for Example 3}
\label{fig_4}
\end{figure}

Consider the same problem of Example 2 with transmission capacity constraints. The results obtained by using SLR are shown in Figure \ref{fig_4}. Near-optimal CH prices with quality of 0.1\% are obtained in 290 sec.  Compared to the duality gap of 2.67\%, the novel quality measure of 0.1\% is more accurate. The amount of time required to obtain the upper bound is 0.7 sec, which is roughly 0.22\% of the total time. The amount of time required to obtain the feasible cost is longer at 1.57 sec. 

Based on the three examples described above, the following observations can be made: 
\begin{enumerate}
    \item The results of Example \ref{ex2} (118-bus without transmission capacity constraints) are comparable to those obtained for a smaller problem of Example \ref{ex1}, in terms of CPU time and CH-price quality. The novel quality measure is significantly more accurate than the duality gap.
    \item The results of Example \ref{ex3} (118-bus with transmission capacity constraints) show the impact of transmission line constraints on solving time and CH-price. However, the novel quality measure is still more accurate than the duality gap.
    \item Compared to the problem-dependent standard duality gaps, which may be large, the novel quality measures are problem-independent and are more accurate.
\end{enumerate}

Computationally, the standard quality gap requires a feasible solution. When the number of units increases, the time required to obtain a feasible solution is expected to increase. The CPU time required to compute the novel quality measure is not expected to increase for larger-scale problems since the number of decision variables within \eqref{eq26} is independent of the number of units, paving the way for providing a tight measure of CH prices for large-scale problems within a modest CPU time. 
\newline

\section{Conclusions }
\label{sec:4}
In this paper, a novel measure for CH prices is developed and the near-optimal CH prices are efficiently obtained by using our recent SLR method. Through numerical testing, it is demonstrated that the near-optimal CH prices with high quality are obtained fast and the computational effort involved in obtaining an upper bound for the CH-price quality measure takes a fraction of the total solving time. The method to quantify the quality of CH prices is generic within the LR-based methods and can be generally used to quantify the quality of the dual solutions. 

\section*{Acknowledgments} 
This work was supported in part by the National Science Foundation under Grants ECCS-1810108, and by a project funded by ISO New England. Any opinions, findings, conclusions or
recommendations expressed in this paper are those of the authors and do not reflect the views of NSF or ISO New England.

 \bibliographystyle{elsarticle-num} 
 
 \bibliography{cas-refs}

%% else use the following coding to input the bibitems directly in the
%% TeX file.

% \begin{thebibliography}{00}

% %% \bibitem{label}
% %% Text of bibliographic item

% \bibitem{}

% \end{thebibliography}
\end{document}